\documentclass{emulateapj}
\usepackage{multirow}
\usepackage{subfigure}

\def\msun{$M_{\odot}$} 
\def\lsun{$L_{\odot}$}

\def\xmm{\textit{XMM-Newton}}
\tabletypesize{\tiny}
\shortauthors{Lin et al.}
\begin{document}

\title{Discovery of the Candidate Off-nuclear Ultrasoft Hyper-luminous X-ray Source 3XMM J141711.1+522541}

 \author{Dacheng Lin\altaffilmark{1}, Eleazar R. Carrasco\altaffilmark{2}, Natalie A. Webb\altaffilmark{3,4}, Jimmy A. Irwin\altaffilmark{5}, Renato Dupke\altaffilmark{5,6,7,8}, Aaron J. Romanowsky\altaffilmark{9,10}, Enrico Ramirez-Ruiz\altaffilmark{11}, Jay Strader\altaffilmark{12}, Jeroen Homan\altaffilmark{13,14},  Didier Barret\altaffilmark{3,4}, Olivier Godet\altaffilmark{3,4} }
\altaffiltext{1}{Space Science Center, University of New Hampshire, Durham, NH 03824, USA, email: dacheng.lin@unh.edu}
\altaffiltext{2}{Gemini Observatory/AURA, Southern Operations Center, Casilla 603, La Serena, Chile}
\altaffiltext{3}{CNRS, IRAP, 9 avenue du Colonel Roche, BP 44346, F-31028 Toulouse Cedex 4, France}
\altaffiltext{4}{Universit\'{e} de Toulouse, UPS-OMP, IRAP, Toulouse, France}
\altaffiltext{5}{Department of Physics and Astronomy, University of Alabama, Box 870324, Tuscaloosa, AL 35487, USA}
\altaffiltext{6}{Observat\'orio Nacional, Rua Gal.\ Jos\'e Cristino 77, S\~ao Crist\'ov\~ao, CEP20921-400 Rio de Janeiro RJ, Brazil}
\altaffiltext{7}{Department of Astronomy, University of Michigan, 500 Church St., Ann Arbor, MI 48109, USA}
\altaffiltext{8}{Eureka Scientific Inc., 2452 Delmer St. Suite 100, Oakland, CA 94602, USA}
\altaffiltext{9}{Department of Physics and Astronomy, San Jos\'{e} State University, One Washington Square, San Jos\'{e}, CA 95192, USA}
\altaffiltext{10}{University of California Observatories, 1156 High Street, Santa Cruz, CA 95064, USA}
\altaffiltext{11}{Department of Astronomy and Astrophysics, University of California, Santa Cruz, CA 95064, USA}
\altaffiltext{12}{Department of Physics and Astronomy, Michigan State University, East Lansing, Michigan, MI 48824, USA}
\altaffiltext{13}{MIT Kavli Institute for Astrophysics and Space Research, MIT, 70 Vassar Street, Cambridge, MA 02139-4307, USA}
\altaffiltext{14}{SRON, Netherlands Institute for Space Research, Sorbonnelaan 2, 3584 CA Utrecht, The Netherlands}

\begin{abstract}
We report the discovery of an off-nuclear ultrasoft hyper-luminous
X-ray source candidate 3XMM~J141711.1+522541 in the inactive S0 galaxy
SDSS~J141711.07+522540.8 ($z=0.41827$, $d_L=2.3$ Gpc) in the Extended
Groth Strip. It is located at a projected offset of $\sim$1\farcs0
(5.2 kpc) from the nucleus of the galaxy and was serendipitously
detected in five \textit{XMM-Newton} observations in 2000 July. Two
observations have enough counts and can be fitted with a standard
thermal disk with an apparent inner disk temperature
$kT_\mathrm{MCD}\sim0.13$ keV and a 0.28--14.2 keV unabsorbed
luminosity $L_\mathrm{X}\sim4\times10^{43}$ erg s$^{-1}$ in the source
rest frame. The source was still detected in three \textit{Chandra}
observations in 2002 August, with similarily ultrasoft but fainter
spectra ($kT_\mathrm{MCD}\sim0.17$ keV,
$L_\mathrm{X}\sim0.5\times10^{43}$ erg~s$^{-1}$). It was not detected
in later observations, including two by \textit{Chandra} in 2005
October, one by \textit{XMM-Newton} in 2014 January, and two by
\textit{Chandra} in 2014 September--October, implying a long-term flux
variation factor of $>$14. Therefore the source could be a transient
with an outburst in 2000--2002. It has a faint optical counterpart
candidate, with apparent magnitudes of $m_\mathrm{F606W}=26.3$ AB mag
and $m_\mathrm{F814W}=25.5$ AB mag in 2004 December (implying an
absolute $V$-band magnitude of $\sim$$-15.9$ AB mag). We discuss
various explanations for the source and find that it is best explained
as a massive black hole (BH) embedded in the nucleus of a possibly
stripped satellite galaxy, with the X-ray outburst due to tidal
disruption of a surrounding star by the BH. The BH mass is $\sim10^5$
\msun, assuming the peak X-ray luminosity at around the Eddington
limit.
\end{abstract}

\section{INTRODUCTION}
\label{sec:intro}
Optical dynamical measurements have confirmed the existence of
stellar-mass black holes (BHs, $\sim$10 $M_\odot$) in some Galactic
X-ray binaries \citep{remc2006,mcre2006} and supermassive BHs (SMBHs;
$\sim$10$^{5-10}$ $M_\odot$) in the centers of massive galaxies
\citep{kori1995}. Most Galactic BH X-ray binaries are discovered when
they experience transient outbursts. Many SMBHs are detected as active
galactic nuclei (AGNs), while quiescent SMBHs can reveal themselves
temporarily through tidal disruption of surrounding stars
\citep{lioz1979,re1988,re1990}. A few tens of tidal disruption event
(TDE) candidates have been discovered in various wavelengths, with
about twenty in X-rays \citep{ko2012,ko2015}. Their positions, at
least for the X-ray candidates, are all consistent with emanating from
the nuclei of their candidate host galaxies.

Off-nuclear/wandering massive BHs, including intermediate-mass BHs
(IMBHs, $\sim$10$^2$--10$^5$ $M_\odot$) and SMBHs, have been predicted
to exist through several important astrophysical processes. The
processes for forming wandering IMBHs include, e.g., the collapse of
massive population III stars in the early Universe
\citep[e.g.,][]{mare2001}; runaway merging of massive stars in young
compact star clusters \citep[e.g.,][]{ebmats2001,gufrra2004}; and
accretion of a large amount of gas lost by the first generation of
giant stars in the center of globular clusters
\citep[e.g.,][]{vemcde2010}. Given that galaxy merging is ubiquitous
and that many dwarf galaxies with optical or X-ray signatures of
nuclear massive BHs have been discovered
\citep{resijo2011,regrge2013,mauler2013,maliir2014,doceco2014,barega2015,lerepl2015},
tidal stripping of merging satellite dwarf galaxies might result in
wandering IMBHs or SMBHs \citep[e.g.][]{istasi2003,begoqu2010}.  A
SMBH of mass $\sim$$2.1\times10^7$ \msun\ has been found in one of the
brightest ultracompact dwarf galaxies (UCDs) yet known, i.e., UCD1 in
M60, and many less massive UCDs were suggested as also ontaining
wandering massive BHs \citep{sevami2014}. Wandering massive BHs
embedded in compact stellar clusters can be revealed in several ways,
such as accreting mass from a stellar companian or tidal disruption of
surrounding stars
\citep{bamaeb2004,hopoal2004,bahopo2006,raro2009,magora2014,matrra2015}.

Many ultra-luminous X-ray sources \citep[ULXs, see][for a
  review]{feso2011}, which are off-nuclear point sources reaching
$L_{\rm X}> 10^{39}$ erg s$^{-1}$ (the Eddington limit for a
stellar-mass BH of $\sim$10 $M_\odot$), have been detected within
nearby galaxies. However, most ULXs are still not luminous enough to
be unambiguously associated with massive BHs; the beaming effect
and/or super-Eddington accretion rates onto stellar-mass BHs can
explain luminosities up to $\sim$$10^{41}$ erg s$^{-1}$
\citep[e.g.,][]{polifa2007}. With dynamical measurements of the masses
of the BHs in two ULXs \citep{librba2013,mopaso2014}, detection of two
transient ULXs in M31 showing transition from the super-Eddington
state to standard spectral states of Galactic BH X-ray binaries
\citep{misuro2012,mimima2013}, and the confirmation of the ubiquitous
curved spectra at high energy in the ULX spectra by \textit{NuSTAR}
\citep[e.g.,][]{barawa2013,wahagr2014}, most ULXs are now believed to
be super-Eddington accreting stellar-mass BHs. \citet{bahawa2014} even
found a ULX (\object{M82 X-2}) powered by an accreting neutron star.

In contrast, hyper-luminous X-ray sources (HLXs, $L_{\rm X}> 10^{41}$
erg s$^{-1}$) are difficult to explain as stellar-mass BHs with
beaming effects and/or super-Eddington accretion rates and are good
targets to search for wandering massive BHs. \object{ESO 243-49 HLX-1}
is the most luminous HLX yet detected, with peak luminosity $L_{\rm
  X}\sim 10^{42}$ erg s$^{-1}$ and thermal disk temperatures of $<$0.3
keV, and has been argued to be an IMBH of $\sim$$10^4$ $M_\odot$
\citep[e.g.,][]{faweba2009,sefali2011,wecsle2012,goplka2012}. There
are a dozen other HLX candidates, which have lower luminosities and
hard X-ray spectra and are mostly persistent with flux variation
factors of a few \citep[e.g.,][]{surowa2012}. Some of them were shown
to be background AGNs \citep[e.g.][]{surowa2012,surogl2015}. Some of
the others show special properties distinguishing them from AGNs and
are good HLX candidates hosting massive BHs (e.g., \object{M82 X-1},
with twin X-ray quasi-periodic oscillations, \citealt{pastmu2014};
\object{CXO J122518.6+144545}, with possible recurrent outbursts,
\citealt{hejoto2015}).

In our continuing effort to classify X-ray sources serendipitously
detected by \textit{XMM-Newton} and \textit{Chandra}
\citep[e.g.,][]{liweba2012,liweba2014}, we discovered a possibly
transient ultrasoft X-ray source \object{3XMM J141711.1+522541}
(XJ1417+52 hereafter) in the \textit{XMM-Newton} Serendipitous Source
Catalog \citep[the 3XMM-DR5 version,][]{rowewa2015}. The source was
serendipitously detected in deep \textit{XMM-Newton} and
\textit{Chandra} observations of the Extended Groth Strip \citep[EGS,
  e.g.,][]{lanage2009} in 2000--2002. We ruled it out as an AGN in
\citet{liweba2012} based on the ultrasoft X-ray spectra. In this
paper, we report the properties of this source and argue that it is
probably a wandering BH of mass $\sim$$10^5$ \msun\ embedded in a
compact stellar cluster at a redshift of $z=0.41827$ (the source
luminosity distance is $d_L=2.3$ Gpc, assuming a flat universe with
$H_0$=70 km~s$^{-1}$~Mpc$^{-1}$ and $\Omega_\mathrm{M}$=0.3), with the
outburst due to disruption of a surrounding star. In
Section~\ref{sec:reduction}, we describe the analysis of
multiwavelength data. In Section~\ref{sec:res}, we first identify the
host galaxy of our source and the optical counterpart, followed by the
presentation of its detailed X-ray spectral and variability
properties. We discuss the nature of our source in
Section~\ref{sec:discussion} and give the conclusions of our study in
Section~\ref{sec:conclusions}.
\section{DATA ANALYSIS}
\label{sec:reduction}
\tabletypesize{\scriptsize}
\setlength{\tabcolsep}{0.02in}
\begin{deluxetable*}{rcccccccc}
\tablecaption{The X-ray Observation Log\label{tbl:obslog}}
\tablewidth{0pt}
\tablehead{\colhead{Obs. ID} &\colhead{Date} & \colhead{Detector} &\colhead{OAA} &\colhead{$T$} &\colhead{$r_\mathrm{src}$}  & \colhead{Count rate} & \colhead{$L_\mathrm{abs}$} & \colhead{$L_\mathrm{unabs}$}\\
 & & & & (ks)& & ($10^{-3}$ counts s$^{-1}$) & \multicolumn{2}{c}{($10^{43}$ erg s$^{-1}$)}\\
(1) & (2) &(3) & (4) & (5) & (6) & (7) & (8) & (9)
}
\startdata
\multicolumn{4}{l}{\xmm:}\\
\hline
0127921001(X1) &2000-07-21 & MOS1/MOS2/pn & 3.4$\arcmin$ & 54.8/54.8/44.2 & 15$\arcsec$/15$\arcsec$/15$\arcsec$ &$2.5\pm0.2$/$1.7\pm0.2$/$4.7\pm0.3$ & $2.3^{+ 0.6}_{-0.4}$ & $3.8^{+ 3.1}_{-1.4}$\\
0127921201(X2) &2000-07-23 & MOS1/MOS2/pn & 3.4$\arcmin$ & 18.3/18.3/12.8 & 15$\arcsec$/15$\arcsec$/15$\arcsec$ & $2.0\pm0.3$/$1.0\pm0.3$/$2.3\pm0.5$ & $1.9^{+  0.9}_{ -0.7}$ & $2.7^{+  8.5}_{ -1.1}$ \\
0723860101(X3) & 2014-01-05 & MOS1/MOS2 & 1.7$\arcmin$  & 26.9/26.8 & 15$\arcsec$/15$\arcsec$ & $<1.0$ & $<0.7$ & $<1.1$ \\
\hline
\multicolumn{4}{l}{\textit{Chandra}:}\\
\hline
3305(C1) & 2002-08-11 & ACIS-I0 &  5.8$\arcmin$ & 29.1 & 4$\farcs$0 & \multirow{3}{*}{$0.14\pm0.03$} & \multirow{3}{*}{$0.5^{+  0.4}_{ -0.3}$} & \multirow{3}{*}{$0.5^{+  0.7}_{ -0.2}$}\\
4357(C2) & 2002-08-12 & ACIS-I0 &  5.8$\arcmin$ & 83.7 & 4$\farcs$0\\
4365(C3) & 2002-08-21 & ACIS-I0 &  5.8$\arcmin$ & 83.6 & 4$\farcs$0\\
5851(C4) & 2005-10-15 & ACIS-I2 & 9.1$\arcmin$ & 35.7 & 7$\farcs$0 & \multirow{2}{*}{$<0.20$} & \multirow{2}{*}{$<0.6$} & \multirow{2}{*}{$<1.0$} \\
7181(C5) & 2005-10-15 & ACIS-I2 & 9.1$\arcmin$ & 16.0 & 7$\farcs$0\\
16027(C6) & 2014-09-15 & ACIS-S3 & 0.2$\arcmin$ & 26.6 & 1$\farcs$2 & \multirow{2}{*}{$<0.10$} & \multirow{2}{*}{$<0.15$} &\multirow{2}{*}{$<0.25$}\\
17487(C7) & 2014-10-11 & ACIS-S3 & 0.2$\arcmin$ & 32.6 & 1$\farcs$2\\
\hline
\enddata 
\tablecomments{Columns: (1) the observation ID with our designation given in parentheses, (2) the observation start date, (3) the detector, (4) the off-axis angle, (5) the exposure times of data used in final analysis, (6) the radius of the source extraction region, (7) the net count rate with $1\sigma$ uncertainties or $3\sigma$ upper limits (0.2--1 keV for \xmm\ observations, 0.3--1 keV for \textit{Chandra} observations C1--C5 and 0.2--1 keV for \textit{Chandra} observations C6--C7, all in the observer frame; the count rate upper limit for X3 is from the combination of MOS1 and MOS2; we note significant loss of counts in MOS2 and pn in X1 and X2 due to the presence of instrumental bad columns), (8) source rest-frame 0.28--14.2 keV luminosity from the MCD fits, corrected for Galactic absorption but not intrinsic absorption, with 90\% uncertainties or $3\sigma$ upper limits, (9) source rest-frame 0.28--14.2 keV luminosity from the MCD fits, corrected for both Galactic and intrinsic absorption, with 90\% uncertainties or $3\sigma$ upper limits. The luminosity upper limits for X3, C4--C5 and C6--C7 were calculated assuming the spectral shape of X1 and have been corrected for PSF loss. All upper limits were calculated with the CIAO task \textit{aprates}, which adopts the Bayesian approach.}
\end{deluxetable*}
\subsection{\textit{XMM-Newton} Observations}
\label{sec:xmmobs}
There were five \textit{XMM-Newton} observations during 2000 July 20--24
that covered XJ1417+52 at off-axis angles of about 3\farcm4. We
analyzed all of them and found that only two have enough counts from
clean exposures for careful spectral modeling; they are referred to as
X1 and X2 (Table~\ref{tbl:obslog}). The other three (observation IDs:
127920401, 0127920901, and 0127921101) have clean exposure times of
$<$10 ks after high background filtering (see below) and have $<$10
net counts in each European Photon Imaging Cameras (EPIC) camera. We
will not present these observations further in this study, but we note
that the source was detected in these observations, with fluxes
consistent with X1 and X2, based on the 3XMM-DR5 catalog. We also performed
an \textit{XMM-Newton} follow-up observation (X3 hereafter, see
Table~\ref{tbl:obslog}) of XJ1417+52 on 2014 January 5. XJ1417+52 was
not clearly detected in this observation, and we use it to constrain
the long-term evolution of the source.

The source was in the field of view (FOV) of all the three EPIC
cameras \citep[i.e., pn, MOS1, and
  MOS2,][]{jalual2001,stbrde2001,tuabar2001} in the imaging mode in
both X1 and X2, but in X3, only MOS1 and MOS2 were collecting science
data.  We used SAS 14.0.0 to analyze the observations. We first
reprocessed the X-ray event files with the calibration files of 2015
July. We excluded the data in strong background flare intervals
following the SAS thread for the filtering against high backgrounds,
i.e., excluding all intervals when the background exceeded the low and
steady
level\footnote{http://xmm.esac.esa.int/sas/current/documentation/threads
  /EPIC\_filterbackground.shtml}. Short flares were seen in X1 in all
cameras, each for $\sim$1\% of the time. Flares were absent in X2 in
all cameras. Short flares also occurred in X3 in both MOS1 and MOS2,
each for 5\% of the time. The final clean exposure times used are
listed in Table~\ref{tbl:obslog}. We extracted the source spectra from
all available cameras using a circular region of radius $15\arcsec$,
corresponding to a point spread function (PSF) enclosing fraction of
$\sim$70\%. The background spectra were extracted from a large
circular region of radius 60$\arcsec$--100$\arcsec$ near the
source. For the event selection criteria, we used the default values
in the pipeline \citep[see Table~5 in][]{wascfy2009}. We also created
the response files, which were used for spectral fits for X1 and X2
and to estimate the flux for X3. To check the short-term variability
in X1 and X2, we extracted the MOS1 background-corrected light curves
in the 0.2--1 keV band (negligible counts above 1 keV) and binned at 6
ks using the SAS tool \textit{epiclccorr}. We note that in both X1 and
X2 our source was coincident with a dark column in MOS2 and a bright
column in pn, resulting in significant loss of counts after filtering
out these columns.

Although the 3XMM-DR5 catalog provides an astrometrically corrected
position for the source using the Sloan Digital Sky Survey (SDSS) DR9
catalog \citep{abadag2009} as the reference, we double-check the
astrometric correction using the deep optical observations of the EGS
by the Advanced Camera for Surveys (ACS) Wide Field Camera (WFC) on
the \textit{Hubble Space Telescope} (\textit{HST}) and
MegaPrime/MegaCam \citep{bochab2003} on the Canada-France-Hawaii
Telescope (CFHT). We started with the source detections in the
3XMM-DR5 catalog for all five observations in 2000 (i.e., including
the three observations not analyzed in detail in this paper). The
source detection in the \emph{XMM-Newton} catalog used a maximum
likelihood fitting procedure \citep{wascfy2009,rowewa2015}. We first
aligned all X-ray observations to the longest observation, i.e., X1,
using the astrometric correction method described in
Appendix~\ref{sec:ascorr} (it is used for all astrometric corrections
throughout the paper), using sources with 95\% positional
uncertainties $<$2\arcsec.  The uncertainties of the X-ray positions
include both the statistical component and a possible systematic
component of 0\farcs37 \citep[$1\sigma$ in R.A. and
  Decl.,][]{rowewa2015}.  After the relative astrometry correction,
the average X-ray source positions weighted by the uncertainties were
obtained and then matched to the optical sources in the
\textit{HST}/ACS/WFC F814W images from the All-Wavelength Extended
Groth Strip International Survey \citep{daguko2007} and the
MegaPrime/MegaCam $i\arcmin$-band images from the CFHT Legacy Survey
\citep[CFHTLS,][]{gw2012}, which were matched to the SDSS
astrometry. Only X-ray sources with 95\% positional uncertainties
$<$1$\arcsec$ were used for the astrometric correction; the
uncertainties of optical sources should be small and were assumed to
be 0\farcs1 ($1\sigma$) in R.A. and Decl. We note that XJ1417+52 was
excluded from the matches used for astrometric correction in this
step, in order to reduce the effect of the astrometric correction on
the identification of its optical counterpart.

\subsection{\textit{Chandra} Observation}
XJ1417+52 was in the FOV of seven \textit{Chandra} observations
(C1--C7 hereafter; refer to Table~\ref{tbl:obslog}), all using the
imaging array of the AXAF CCD Imaging Spectrometer \citep[ACIS;
][]{bapiba1998}. C1--C5 were part of the \textit{Chandra} survey of
the EGS \citep{lanage2009}, with C1--C3 taken between 2002 August
11--21 and C4--C5 taken on 2005 October 15. Our source fell in the
front-illuminated chip I0 in C1--C3 with an off-axis angle of 5\farcm8
and in I2 in C4--C5 with an off-axis angle of 9\farcm1 (near the CCD
edge). C6 and C7 were our follow-up observations of the source in 2014
September--October. The aim point was chosen to be at the
back-illuminated chip S3 because our source had been ultrasoft.  We
applied the latest calibration (CALDB 4.6.7) by reprocessing all the
data with the script \textit{chandra\_repro} in the \textit{Chandra}
Interactive Analysis of Observations (CIAO, version 4.7) package. We
extracted source and background spectra from circular regions and
created the corresponding response matrices for all observations using
the script \textit{specextract}. The radii of the source regions in
C1--C3, C4--C5, and C6--C7 are $4\farcs0$, $7\farcs0$, and $1\farcs2$
(Table~\ref{tbl:obslog}), corresponding to PSF enclosing fractions of
90\%, 70\%, and 95\% at 0.6 keV, respectively. We used a source region
of a smaller PSF enclosing fraction for observations in which the
source has a larger off-axis angle, in order to reduce the background
effect, because the PSF of \textit{Chandra} degrades significantly at
large off-axis angles. A large background region of radius $30\arcsec$
was used for all observations. Our source was detected in C1--C3, but
not in C4--C7. We used \textit{combine\_spectra} to combine the
spectra of C1--C3 into a single spectrum for spectral fitting, because
these observations are close in time and have very few source counts
(the total net source counts are 38 in 0.3--7.0 keV). Similarly we
combined the spectra of C4 and C5 into a single spectrum and the
spectra of C6 and C7 into another one. These combinations were to put
a tighter constraint on the limit of the source flux in these
observations, in which our source was not detected.

In order to determine the position of XJ1417+52 from C1--C3, in which
the source was weakly detected, we merged these observations after
correcting the relative astrometry between them. The astrometric
correction used sources detected with the CIAO wavelet-based source
detection tool {\it wavdetect} \citep{frkaro2002}, which was applied
to 0.5--7.0 keV count images binned at the single sky pixel resolution
(0\farcs492). The exposure maps were constructed at the corresponding
monochromatic effective energy, i.e., 2.3 keV \citep{evprgl2010}. The
PSF maps used correspond to the 40\% enclosed counts fraction at 2.3
keV. Only sources with 95\% positional uncertainties \citep[based on
  Equation 12 of][]{kikiwi2007} $<$2\farcs0 were used for astrometric
correction. C2 was used as the astrometric reference. New aspect
solution files were created from the relative astrometric correction
obtained and then applied to the event files for C1 and C3. The CIAO
script {\it merge\_obs} was then used to combine the event lists of
C1--C3. The source detection was then carried out on the merged
observation twice. The first time was on the 0.5--7.0 keV count image
to detect sources and align the astrometry to that of optical sources
from \textit{HST}/ACS/WFC F814W-band and CFHTLS MegaPrime/MegaCam
$i\arcmin$-band images, as we did to obtain \textit{XMM-Newton}
position of our source. The second time was on the 0.3--1.0 keV count
image (the exposure and PSF maps at 0.6 keV were used) to obtain the
position of XJ1417+52, considering that our source was ultrasoft. In
order to calculate the statistical positional uncertainty for our
source, we carried out 2000 ray-trace simulations with {\it MARX}
5.1.0 at positions near it and at the same off-axis angle. The
spectrum from the multicolor disk (MCD) fit to C1--C3
(Section~\ref{sec:spmodel}) was assumed.

\subsection{The \textit{Hubble} Space Telescope Images}
There are two images (in two filters: F606W and F814W) taken by the
\textit{HST}/ACS/WFC on 2004 December 15 in the field of our X-ray
source. Each image has four frames of 525 seconds each. As will be
shown later in Section~\ref{sec:multiwav}, these images indicate that
our source has a candidate host galaxy
\object{SDSS~J141711.07+522540.8} (GJ1417+52 hereafter) with extent
$\sim$4\arcsec\ and a very faint candidate optical counterpart
(sGJ1417+52 hereafter) at a $\sim$1\arcsec\ offset from the
nucleus. In order to derive their main photometric parameters, we
fitted the \textit{HST}/ACS/WFC images using GALFIT \citep{pehoim2010}
with multiple S\'{e}rsic components (convolved with the PSF) for the
galaxy and a PSF model for the faint optical source, which seems
unresolved in these images. In order to improve the statistics,
especially for the faint source sGJ1417+5, the fits used
$5\arcsec\times5\arcsec$ (centered at the center of the galaxy)
stacked images (one for each filter) from aligned and
bundled\footnote{http://www.stsci.edu/hst/wfc3/documents/ISRs/WFC3-2014-24.pdf}
FLC frames. The effective
PSFs\footnote{http://www.stsci.edu/hst/acs/documents/isrs/isr0601.pdf}
at the position of our source in the four frames were averaged and
used to fit the stacked images. Because of dithering in the
observations, different frames had different degrees of distortion at
the position of our source, causing some problem in stacking
them. However, we found that the fits to single frames with the
corresponding PSFs gave similar results.

\begin{figure*}[!tb]
  \centering
\includegraphics[width=6.4in]{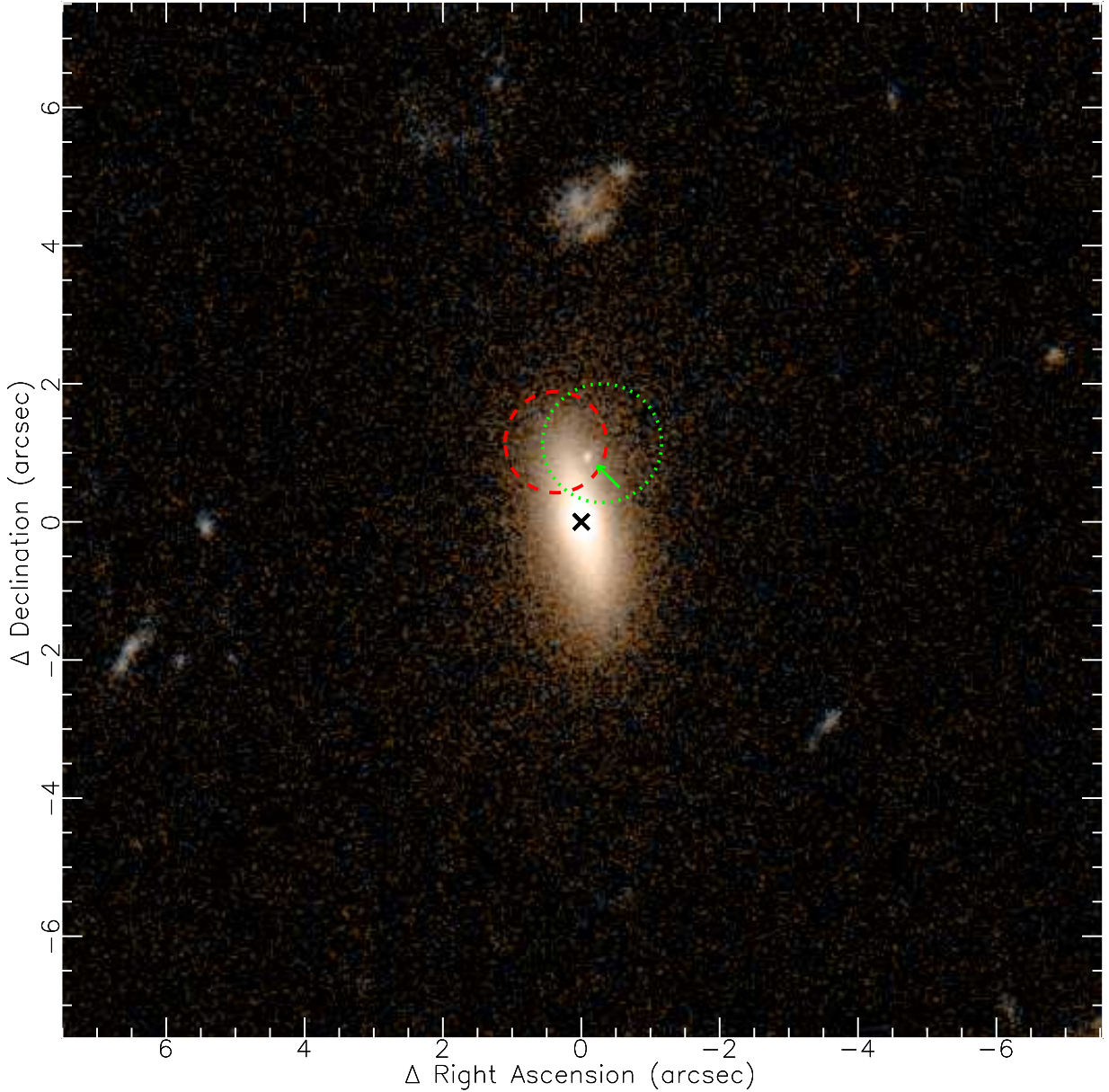}
\caption{{\it HST}/ACS/WFC multidrizzled image in the field of
  XJ1417+52, with the origin at the center of the galaxy GJ1417+52
  (black cross, R.A.= 14:17:11.076, Decl.=$+$52:25:40.80). The angular
  scale of GJ1417+52 is 5.5 kpc/1$\arcsec$. The image is
  false-colored, using the F814W (red) and F608W (blue) images and
  their mean (green). The green arrow points to a faint optical source
  sGJ1417+52 at R.A.=14:17:11.066 and Decl.=$+$52:25:41.74. The 95\%
  positional error (0\farcs73) of XJ1417+52 from the
  \textit{XMM-Newton} observations is marked as a red dashed circle,
  and that (0\farcs86) from the {\it Chandra} observations as a green
  dotted circle, both indicating that the faint optical source could
  be the counterpart to our X-ray source.
  \label{fig:s132740hstimg}}
\end{figure*}

\subsection{The Gemini Spectroscopic Observation}
The galaxy GJ1417+52 was observed during the night of 2013 February 8
(UT) with the Gemini Multi-Object Spectrograph
\citep[GMOS,][]{hojoal2004} at the Gemini North Telescope, in queue
mode (program ID GN-2013A-Q-37). The data were acquired during dark
time (illumination fraction 0.8\%), under photometric conditions and
$\sim$0\farcs70 seeing. The 400 lines/mm ruling density grating (R400)
centered at 7000 \AA\ was used. We chose a slit of width 0\farcs75 and
put it through the center of the galaxy GJ1417+52 and the source
sGJ1417+52 in order to obtain their spectra simultaneously. A total of
four exposures of 1500 seconds each were obtained. Small offsets in
the spectral direction (50 \AA) towards the blue and the red were
applied between exposures to allow for the gaps between CCDs to avoid
any loss of important lines present in the spectra.  Spectroscopic
flats and comparison lamp (CuAr) spectra were taken after each science
exposure.  In addition, to derive the sensitivity function and flux
calibrate the science spectrum, the spectrophotometric standard star
G191B2B was observed as part of the baseline calibration provided by
the observation. Because the standard star was observed on a different
night (2013 March 03 UT) and under different observing conditions, the
science spectrum was calibrated in relative flux.

We processed the observations and extracted the spectrum for
the galaxy GJ1417+52 following the standard procedures for
long-slit observations provided by the Gemini/GMOS package in IRAF. In
summary, the science exposures, comparison lamps and spectroscopic
flats were bias subtracted and trimmed. Spectroscopic flats were then
processed to remove the calibration unit plus GMOS spectral response,
normalizing and leaving only pixel-to-pixel variations and
fringing. The bias subtracted, flat fielded two-dimensional science
spectra were then wavelength calibrated and rectified (S-shape
distortions removed), sky-subtracted, extracted to a one-dimensional
format using a fixed aperture of 1\farcs2 in width around the center
of the galaxy, and average combined.

The final spectrum of the galaxy has a resolution of $\sim$5.5
\AA\ FWHM (measured using sky lines at $\sim$7000 \AA), a dispersion
of $\sim$1.36 \AA\ pixel$^{-1}$, and a signal-to-noise ratio about 40
at 7000 \AA, covering a wavelength interval from $\sim$4850 \AA\ to
9180 \AA. We identified the most prominent absorption lines (as no
clear emission lines were detected) in the spectrum and derived the
redshift by employing a line-by-line Gaussian fit using the {\it
  rvidline} routine in the IRAF RV package.  We fitted the spectrum to
multi-component models comprised of single-population synthetic
spectra, using Penalized Pixel Fitting ({\tt pPXF}) software
\citep{caem2004} and \cite{vasafa2010} synthetic spectra spanning a
grid of 24 ages between 1 to 14 Gyr and 6 metallicities
[M/H]=\{$-1.71$, $-1.31$, $-0.71$, $-0.40$, $0.00$, $+0.22$\}.

The source sGJ1417+52 is very faint and has strong contamination from
the extended emission of the galaxy, and our Gemini observation is
only useful for searching for strong emission lines from it. We
followed a method similar to \citet{sokuwi2013} for \object{ESO 243-49
  HLX-1} to obtain a galaxy subtracted spectrum for this source. We
first aligned and stacked the sky background subtracted 2D spectra
from the four exposures. The galaxy emission was then modeled and
subtracted by fitting a third-order cubic spline function along the
spatial direction on the northern half of the galaxy, excluding 6
pixels ($\sim$0\farcs9) centered around sGJ1417+52.

\section{RESULTS}
\label{sec:res}
\subsection{The Source Position and the Environment}
\label{sec:multiwav}

The positions of XJ1417+52 that we obtained from the
\textit{XMM-Newton} and \textit{Chandra} observations are
R.A.=14:17:11.11, Decl.=+52:25:42.0, and (R.A.=14:17:11.04,
Decl.+52:25:41.9, with the 95\% uncertainties of 0\farcs73 and
0\farcs86, respectively. They are separated by 0\farcs67 but are
consistent with each other within the uncertainties. The
\textit{XMM-Newton} position that we obtained is only 0\farcs16 from
that given in the 3XMM-DR5 catalog, and thus they are also consistent
with each other within the uncertainties. We show the {\it
  HST}/ACS/WFC F606W and F814W false-colored image in the field of
XJ1417+52 in Figure~\ref{fig:s132740hstimg}, with the X-ray positions
above denoted. Our source is close to GJ1417+52, which seems to be an
S0 galaxy. However, neither of the X-ray positions is consistent with
the galaxy center within the 95\% uncertainties. Instead, both X-ray
positions are consistent with a faint but clearly visible optical
source (i.e., sGJ1417+52, pointed to by a green arrow in
Figure~\ref{fig:s132740hstimg}) in the northern part of the galaxy
within the 95\% uncertainties. Based on the \textit{HST} observations
of the EGS, we calculated the chance probability for our X-ray source
to be within $\sim$1\farcs0 of the center of a galaxy similar to or
brighter than GJ1417+52 in the F814W band to be 0.03\%. Similarly,
based on the density of optical sources as bright as or brighter than
sGJ1417+52 in the F814W band, we calculated the chance probability for
sGJ1417+52 to be within $\sim$1\farcs0 of our X-ray source is
5\%. These probabilities are relatively low and allow us to conclude
that XJ1417+52 is most likely in GJ1417+52, with sGJ1417+52 being the
optical counterpart.

The results of our fits to the {\it HST}/ACS/WFC images of GJ1417+52
and sGJ1417+52 are given in Table~\ref{tbl:hstfit}. We required three
S\'{e}rsic components to fit the galaxy, and adding another component
did not improve the fit significantly. The fits are good, with no
clear large residuals left. We added a PSF model in the fits to check
whether a bright point source was present at the galaxy center but saw
no significant improvement on the fits either; the central point
source, if present, would be $\sim$4 mag fainter than the
galaxy. Based on the sizes, indices, and axis ratios, the first and
second S\'{e}rsic components in Table~\ref{tbl:hstfit} are probably
the bulge and the disk, respectively. The third one is much larger in
size (effective radius $R_\mathrm{e}\sim8$ kpc) and could be a
halo. The galaxy has integrated magnitudes of $m_{F606W}=20.28$ AB mag
and $m_{F814W}=19.35$ AB mag, and the optical source sGJ1417+52 has
$m_{F606W}=26.33$ AB mag and $m_{F814W}=25.51$ AB mag, thus about 6
mag fainter than the galaxy. To put a constraint on the size of
sGJ1417+52, we tried to model it with a S\'{e}rsic profile (convolved
with the PSF) instead of a single PSF. We assumed an axis ratio of 1.0
and considered two possible indices: $n=1.0$ and 4.0. The $3\sigma$
upper limits of the effective radius $R_\mathrm{e}$ were found to be
63 pc and 113 pc for the F606W and F814W bands, respectively, in the
case of $n=1.0$, and were 30 pc and 80 pc, respectively, in the case
of $n=4.0$. The best-fitting $R_\mathrm{e}$ values were consistent
with zero (i.e., reduced to be a PSF) within 1$\sigma$ in all cases.

\begin{figure*}
\centering
\includegraphics[width=0.95\textwidth]{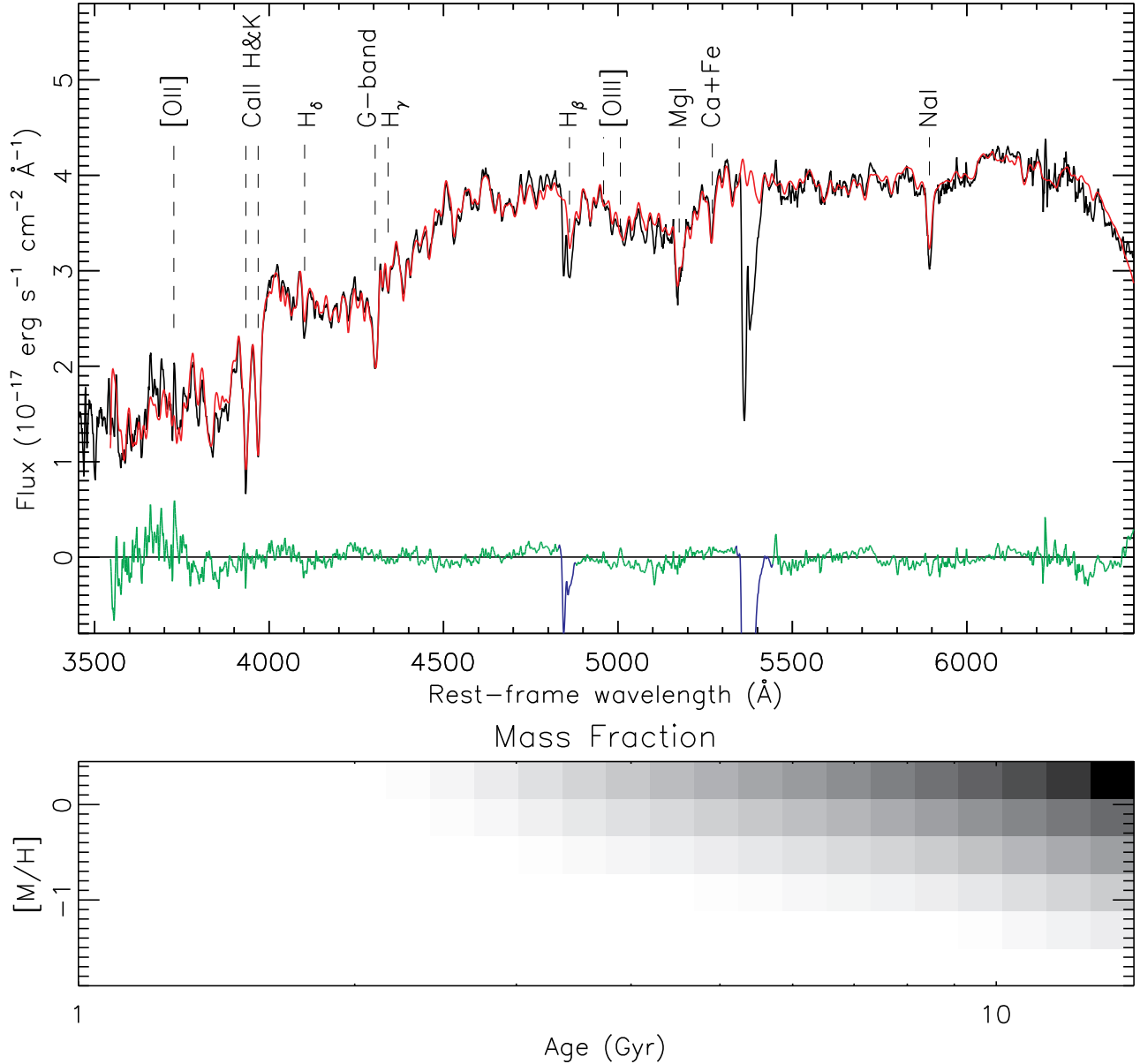}
\caption{(Upper panel) Relative flux calibrated Gemini spectrum
  (black) of the host galaxy of XJ1417+52 versus the source rest-frame
  wavelength, with the best-fitting pPXF model (red) and residuals
  (green/blue) overplotted. The two drops at around 4844 \AA\ and
  5360 \AA\ (corresponding to 6870 \AA\ and 7604 \AA, respectively, in
  the observer frame; indicated by the blue residuals), are due to the
  atmospheric OH absorption and were excluded in the fit. The spectrum
  bluer than 3541 \AA\ was not fitted because the stellar models do not
  cover this wavelength range. Important stellar absorption and AGN
  diagnostic emission lines are denoted for reference. For clarity, we
  have smoothed data points with a boxbar function with width 5 pixels
  for clarity. (Lower panel) Relative mass fractions of different
  stellar populations with respect to metallicity and age, with darker
  shading indicating a larger mass fraction in the best-fitting
  model. }
  \label{fig:s132740gsp}
\end{figure*}

\begin{deluxetable*}{lccccc}
\tablecaption{Fitting results of the \textit{HST}/ACS/WFC images around the field of XJ1417+52\label{tbl:hstfit}}
\tablewidth{0pt}
\tablehead{Components\tablenotemark{a} & 1st S\'{e}rsic & 2nd S\'{e}rsic & 3rd S\'{e}rsic & PSF}
\startdata
F606W \\
Integrated magnitude (AB mag) &$21.57\pm0.01$ &  $21.27\pm0.03$ &  $21.62\pm0.03$ & $26.33\pm0.04$ \\
Effective radius (pixel\tablenotemark{b}) & $1.63\pm0.02$ &  $14.02\pm0.17$ &  $30.97\pm0.62$ & \nodata \\
Index & $1.78\pm0.04$ &  $1.01\pm0.02$ &  $0.76\pm0.04$ & \nodata \\
Axis ratio & $0.65\pm0.01$ &  $0.25\pm0.01$ &  $0.52\pm0.01$ & \nodata \\
Position angle (degree)\tablenotemark{c} & $14.95\pm0.64$ &  $13.01\pm0.11$ &  $3.45\pm1.11$ & \nodata \\
\hline
F814W \\
Integrated magnitude (AB mag) &$20.49\pm0.01$ &  $20.48\pm0.02$ &  $20.66\pm0.02$ & $25.51\pm0.03$ \\
Effective radius (pixel\tablenotemark{b}) & $1.69\pm0.02$ &  $13.48\pm0.12$ &  $28.83\pm0.34$ & \nodata \\
Index & $1.71\pm0.03$ &  $0.79\pm0.02$ &  $0.65\pm0.02$ & \nodata \\
Axis ratio & $0.58\pm0.01$ &  $0.26\pm0.01$ &  $0.47\pm0.01$ & \nodata \\
Position angle (degree)\tablenotemark{c} & $15.45\pm0.41$ &  $13.01\pm0.08$ &  $6.17\pm0.47$ & \nodata 
\enddata 
\tablecomments{\textsuperscript{a}The three S\'{e}rsic components were used to fit the galaxy GJ1417+52 (their centers were consistent with each other and were thus forced to be the same in the final fits), and the PSF was used to fit the faint optical source sGJ1417+52. \textsuperscript{b}The pixel scale is 0\farcs05 (i.e., 0.28 kpc). \textsuperscript{c}The position angle is from the north to the east.}
\end{deluxetable*}

The Gemini spectrum of the galaxy GJ1417+52 is shown in
Figure~\ref{fig:s132740gsp}. The spectrum exhibits no clear emission
lines, but has typical absorption features indicating a passive galaxy
at a redshift of $z=0.41827\pm0.00011$ ($D_L=2.3$ Gpc). We estimated
the $3\sigma$ upper limit of the flux of [\ion{O}{3}] $\lambda$5007 to
be $3.6\times10^{-18}$ erg s$^{-1}$ cm$^{-2}$, which corresponds to a
luminosity of $2.3\times10^{39}$ erg s$^{-1}$ after Galactic
extinction correction. Applying the bolometric correction factors from
the [\ion{O}{3}] $\lambda$5007 flux in \citet{labima2009}, we obtained
the $3\sigma$ upper limit of the bolometric luminosity of the
persistent nuclear activity to be $2.0\times10^{41}$ erg s$^{-1}$. We
note that the MMT had spectroscopic follow-up on the X-ray sources in
the EGS in 2007--2008, with GJ1417+52 being one of the targets. The
spectrum has much lower quality than the Gemini one, with the upper
limit of the flux of [\ion{O}{3}] $\lambda$5007 estimated to be much
higher \citep[$4.8\times10^{-17}$ erg s$^{-1}$ cm$^{-2}$,
  $2\sigma$,][]{yahone2011}, but \citet{cogene2009} obtained a
redshift ($z=0.4184$) consistent with our results.

The pPXF fitting results are shown in Figure~\ref{fig:s132740gsp}. The
light-weighted age is 8.4 Gyr, while the mass-weighted age is 9.4
Gyr. The total mass is $\sim$$4.1\times10^{11}$ \msun, and the total
luminosity within the fitting band (source rest-frame 3541-6479 \AA)
is $\sim$$2.8\times10^{10}$ \lsun, after rescaling the spectrum to
match the integrated F814W flux to correct for the slit loss and the
calibration uncertainty.

The fit implied a stellar velocity dispersion of
$\sigma_\star\sim247\;\rm{km\;s}^{-1}$. Using the relation between
$M_\mathrm{BH}$ and $\sigma_\star$ in \citet{gurige2009}, we inferred
$M_\mathrm{BH}\sim4.0\times10^8\;$\msun\ (the $1\sigma$ uncertainty is
0.31 dex).  We also estimated the central BH mass of the galaxy based
on the BH mass versus bulge rest-frame $K$-band luminosity relation
\citep{gr2007,mahu2003}. The $K$ band in the source rest frame is
approximately the Wide-field Infrared Survey Explorer (WISE) $W1$ band
\citep{wreima2010} in the observer frame for GJ1417+52, which has a
magnitude of $m_{W1}=15.7$ mag ($M_{W1}=-25.7$ mag). Thus we
alternatively estimated the BH mass to be $\sim3.1\times10^8$
\msun\ (the $1\sigma$ uncertainty is 0.33 dex), assuming the
bulge-to-total luminosity ratio of 35\% obtained from the fits to the
\textit{HST} F814W image above. The above two estimates of the central
BH mass agree with each other very well.

At the redshift of GJ1417+52, the source sGJ1417+52 has a projected
offset of $\sim$5.2 kpc from the nucleus of the galaxy. We did not
find any continuum emission, as expected considering that it is so
faint, or any significant emission line from its galaxy subtracted 2D
spectrum at any wavelength covered by our Gemini observation
($\sim$4850 \AA\ to 9180 \AA). We fitted its \textit{HST} F606W and
F814W photometry with the \citet{ma2005} stellar population model that
is based on theoretical atmospheres with the Salpeter initial mass
function. We adopted this model considering its broad wavelength
coverage as needed here. We assumed a single population with a solar
metallicity and a Galactic reddening value of $E(B-V)=0.039$ mag
\citep[][the intrinsic reddening is neglected considering that the
  absorption inferred from X-ray spectral fits is consistent with
  zero]{scfida1998}. The redshift of $z=0.41827$ was applied. We
inferred a stellar population of age 0.8 Gyr and bolometric luminosity
$2.3\times10^{42}$ erg s$^{-1}$ (or $5.9\times10^8$ \lsun,
corresponding to $5.9\times10^7$ \msun). If we assume a systematic
error of 0.1 mag in our derivation of the photometry, the 90\% upper
limit of the age would be 3 Gyr. Assuming at this age, the bolometric
luminosity would be $2.4\times10^{42}$ erg s$^{-1}$ (or
$6.4\times10^8$ \lsun, corresponding to $7.8\times10^8$ \msun). With
photometric information in two filters only, we cannot test multiple
stellar population models. Considering that sGJ1417+52 has a
F606W$-$F814W color similar to GJ1417+52, we cannot rule out that
sGJ1417+52 contains multiple stellar populations, with mass dominated
by old populations.

\subsection{X-ray Spectral Modeling}
\label{sec:spmodel}
\begin{figure}[!tb]
\centering
\includegraphics[width=3.4in]{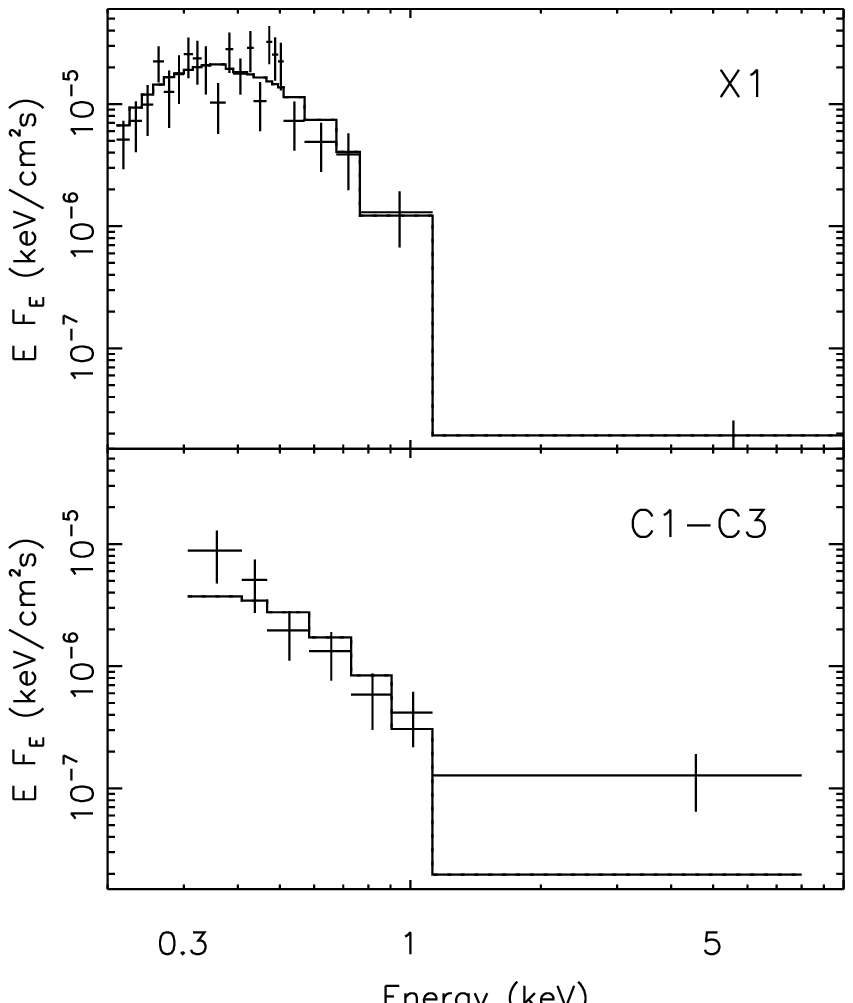}
\caption{The unfolded spectra of X1 (upper panel, for clarity, only
  the MOS1 spectrum is shown) and C1--C3 (lower panel) from the MCD fits. For clarity, the spectra were rebinned to be
  above 2$\sigma$ per bin in the plot. \label{fig:srcspec}}
\end{figure}

We carried out spectral fits for X1, X2 and C1--C3. Because of their
poor statistics, we rebinned the source spectra to have a minimum of
one count per bin and adopted the C statistic to fit the source and
background spectra simultaneously. We fitted over the 0.2--10 keV and
0.3--8 keV energy bands for \textit{XMM-Newton} and \textit{Chandra}
spectra, respectively. As our source is most likely associated with
GJ1417+52 at $z=0.41827$ (Section~\ref{sec:multiwav}), instead of
being a foreground source (see discussion in
Section~\ref{sec:discussion}), we applied this redshift to the
spectral models using the convolution model \textit{zashift} in XSPEC,
unless indicated otherwise. All models included the Galactic
absorption of $N_\mathrm{H}=1.1\times10^{20}$ cm$^{-2}$
\citep{kabuha2005} using the \textit{tbabs} model. Possible absorption
intrinsic to the source was also taken into account using the
\textit{ztbabs} model. We used the abundance tables of
\citet{wialmc2000}. The uncertainties of the parameters from the
spectral fits are all at the 90\% confidence level throughout the
paper.

The X-ray spectra are ultrasoft, with little emission above 1
keV. When we fitted the spectra with a power law (PL), we obtained
unphysically high photon indices of $\Gamma_\mathrm{PL}=7.5\pm1.2$,
$7.0\pm2.1$, and $5.3^{+2.2}_{-0.7}$, respectively, implying the
thermal origin of the spectra. Therefore, we fitted the spectra with a
single-temperature blackbody (BB, \textit{bbodyrad} in XSPEC) model
and an MCD model (\textit{diskbb} in XSPEC). The fitting results are
given in Table~\ref{tbl:spfit}, and the example MCD fits to X1 and
C1--C3 are shown in Figure~\ref{fig:srcspec}. The source rest-frame
temperatures are in the range of $kT_\mathrm{BB}\sim0.11$--0.14 keV
from the BB fits and in the range of $kT_\mathrm{MCD}\sim0.13$--0.17
keV from the MCD fits. Although it seems that the best-fitting
temperature is a little lower in X1 than in C1--C3 (e.g.,
$kT_\mathrm{MCD}=0.13\pm0.02$ keV versus $0.17\pm0.04$ keV), they are
consistent within the 90\% uncertainties. The slightly higher
best-fitting temperature in C1--C3 from the BB and MCD fits could be
due to presence of some very weak hard emission above 1 keV in these
observations. When we tried to fit X1 and C1--C3 spectra with an MCD
plus a PL, with the photon index fixed at 2.0, we obtained a zero PL
normalization for X1 and a non-zero PL normalization for C1--C3 but
only at the 90\% significance level. The best-fitting disk temperature
becomes $kT_\mathrm{MCD}=0.15\pm0.04$ keV for C1--C3, thus closer to
that of X1. The strength of the soft excess $R_\mathrm{exc}$, measured
by the ratio of unabsorbed 0.3--2 keV (source rest-frame) flux in the
MCD and PL components, is $>$61 (the 90\% lower limit) and $\sim$34
for X1 and C1--C3, respectively, which are much higher than those of
ultrasoft AGNs \citep[$R_\mathrm{exc}\lesssim17$,][]{gido2004}. We
will describe the source luminosities from these spectral fits in
Section~\ref{sec:var}.

For easy comparison with Galactic sources, we also fitted the spectra
assuming the X-ray source to be in our Galaxy. Adopting an absorbed BB
model, we obtained $kT_\mathrm{BB}=79\pm8$ eV, $85\pm14$ eV, and
$99\pm18$ eV, for X1, X2, and C1--C3, respectively. The corresponding
0.3--10 keV unabsorbed luminosities are
$2.8_{-0.6}^{+1.1}\times10^{30}d^2$ erg s$^{-1}$,
$2.2_{-0.6}^{+2.0}\times10^{30}d^2$ erg s$^{-1}$, and
$5.2_{-1.7}^{+4.3}\times10^{29}d^2$ erg s$^{-1}$, where $d$ is the
source distance in units of kpc, respectively. Therefore the source
would be very faint if it is in our Galaxy. The best-fitting column
densities are $N_\mathrm{H}=1.0^{+2.2}\times10^{20}$ cm$^{-2}$,
$0.5^{+4.7}\times10^{20}$ cm$^{-2}$, and $0.0^{+6.2}\times10^{20}$
cm$^{-2}$ (the lower error bounds of $N_\mathrm{H}$ are all zeros),
respectively.

\subsection{The Long-term and Short-term X-ray Variability}
\label{sec:var}
\begin{figure}[!tb]
\centering
\includegraphics[width=3.4in]{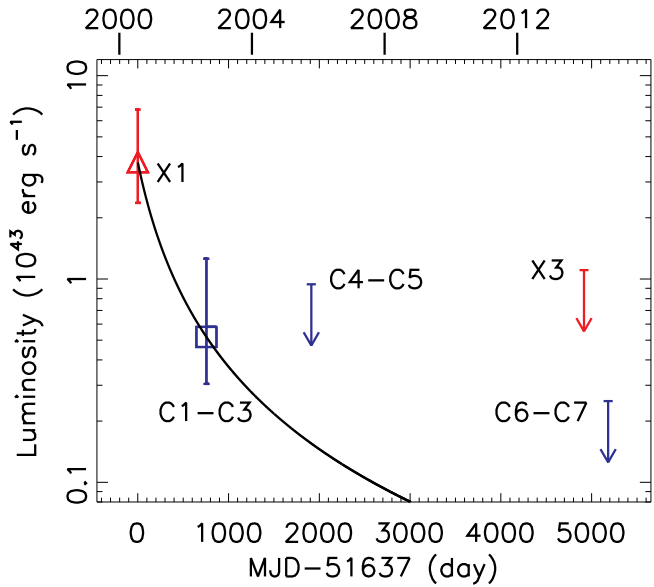}
\caption{The long-term rest-frame 0.28--14.2 keV unabsorbed luminosity
  curve, with 90\% uncertainties or $3\sigma$ upper limits (see
  Table~\ref{tbl:obslog}). For clarity, X2 is not plotted because it
  is only two days after X1 and had a similar luminosity (but with a
  larger uncertainty). The solid line represents a
  $(t-t_\mathrm{D})^{-5/3}$ decline passing through the X1 and C1--C3
  data points, which implies the disruption time $t_\mathrm{D}$ to be
  $\sim$11 months before X1. \label{fig:lumlc}}
\end{figure}

\begin{figure*}[!tb]
\centering
\includegraphics[width=6.8in]{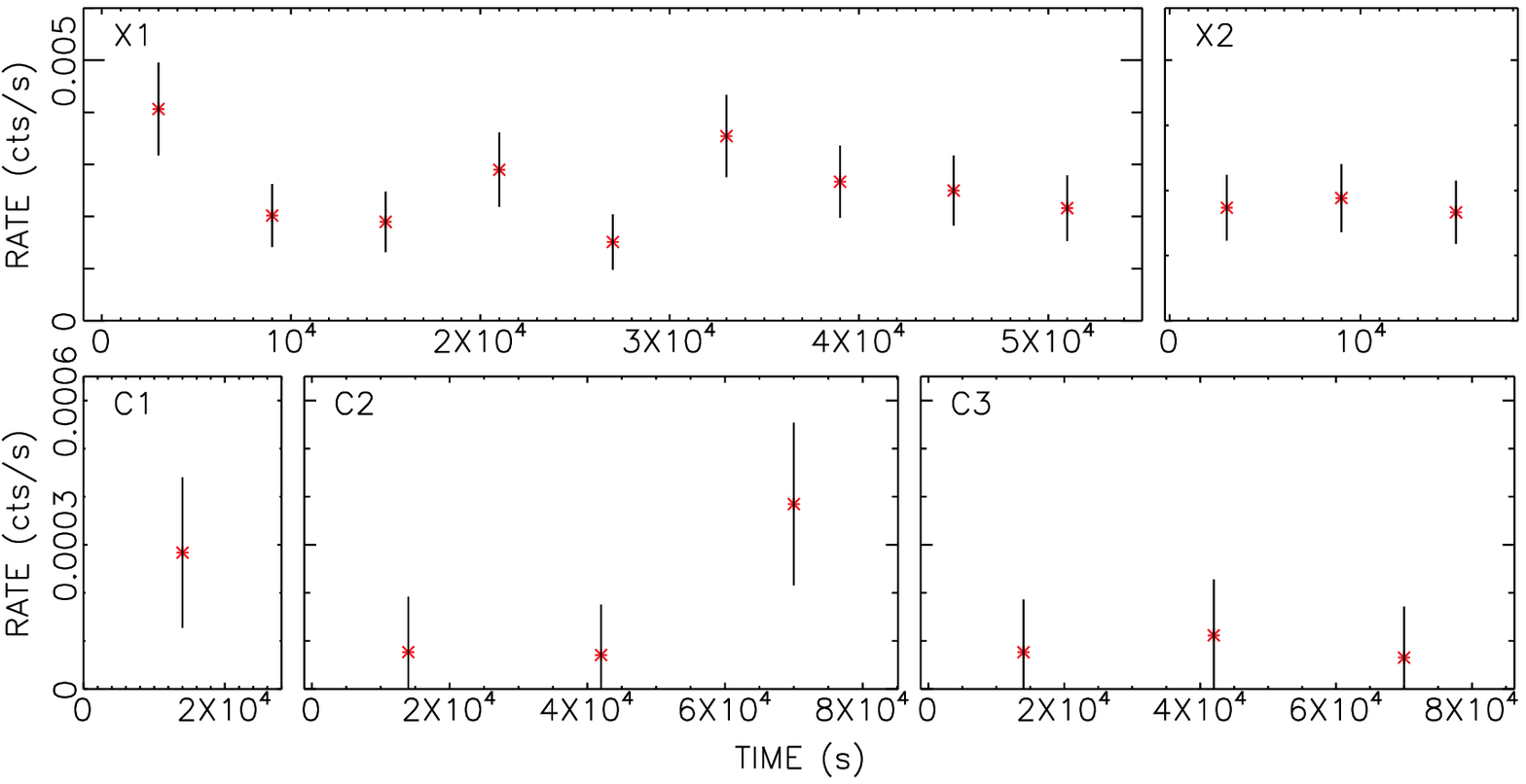}
\caption{The background subtracted light curves for
  \textit{XMM-Newton} observations X1 and X2 (upper panels, MOS1
  camera, 0.2--1 keV, bin size 6 ks) and for \textit{Chandra}
  observations C1--C3 (lower panels, 0.3--1 keV, bin size 28 ks). We
  note that X1 and X2 are separated by two days, while C2 is one day
  after C1 and C3 is nine days after C2. The $1\sigma$ uncertainties
  are included, and they are calculated following \citet{ge1986} for
  the \textit{Chandra} observations due to low counts of most data
  points.  \label{fig:srclc}}
\end{figure*}

Figure~\ref{fig:lumlc} shows the long-term rest-frame 0.28--14.2 keV
(observer-frame 0.2--10 keV) unabsorbed luminosity $L_\mathrm{X}$
curve of XJ1417+52. The luminosities were obtained based on the MCD
fits (Section~\ref{sec:spmodel}) and assuming $D_L=2.3$ Gpc
(Section~\ref{sec:multiwav}); for observations in which the source was
not detected, the $3\sigma$ upper limits were estimated based on the
MCD fit to the X1 spectrum. The source was first detected in X1 and X2
in 2000 July, with $L_\mathrm{X}=3.8^{+ 3.1}_{-1.4}\times10^{43}$
erg~s$^{-1}$ and $2.7^{+ 8.5}_{-1.1}\times10^{43}$ erg~s$^{-1}$,
respectively. The source was still detected in C1--C3 in 2002 August,
with $L_\mathrm{X}=0.5^{+ 0.7}_{ -0.2}\times10^{43}$ erg~s$^{-1}$, a
factor of $\sim7$ lower than that in X1. The source was not detected
in C4--C5, X3, and C6--C7, with $L_\mathrm{X}<0.9\times10^{43}$
erg~s$^{-1}$, $<1.1\times10^{43}$ erg~s$^{-1}$, and
$<0.25\times10^{43}$ erg~s$^{-1}$, respectively. Therefore it appears
that the source was experiencing an outburst in 2000--2002, with the
X-ray luminosity decreasing by a factor of $>$14 in C6--C7 from X1.

Figure~\ref{fig:srclc} shows the light curves from X1, X2, and
C1--C3. The temporal bin sizes used are relatively large (6 ks for X1
and X2 and 28 ks for C1--C3) due to the poor statistics of all
observations. Short-term variability might be present in X1 and C2 but
is not significant. The probability that the source is not variable is
52\% for X1.
\begin{deluxetable}{lccc}
\tablecaption{Fitting results of the X1, X2 and C1--C3 spectra of XJ1417+52\label{tbl:spfit}}
\tablewidth{0pt}
\startdata
\hline
X1 \\
Models & BB & MCD\\
$N_\mathrm{H,i}$ (10$^{20}$ cm$^{-2}$) & $0.0^{+7.4}$ & $  4.5^{+7.6}$\\
$kT_\mathrm{MCD}$/$kT_\mathrm{BB}$ (keV)& $0.113^{+ 0.007}_{-0.014}$  & $0.132^{+ 0.019}_{-0.019}$\\
$N_\mathrm{MCD}$/$N_\mathrm{BB}$ &$ 60^{+127}_{-19}$ & $43^{+124}_{-28}$\\
$L_\mathrm{abs}$ (10$^{43}$ erg~s$^{-1}$)\tablenotemark{a}  & $ 2.4^{+ 0.4}_{-0.5}$  & $ 2.3^{+ 0.6}_{-0.4}$ \\
$L_\mathrm{unabs}$ (10$^{43}$ erg~s$^{-1}$)\tablenotemark{b} & $ 2.4^{+ 1.8}_{-0.4}$  & $ 3.8^{+ 3.1}_{-1.4}$ \\
$L_\mathrm{bol}$ (10$^{43}$ erg~s$^{-1}$)\tablenotemark{c} & $ 3.4^{+ 3.2}_{-0.6}$  & $ 8.9^{+10.7}_{-3.9}$\\
$C/\nu$\tablenotemark{d} & $198.6(194)$ & $197.1(194)$\\
\hline
X2 \\
Models & BB & MCD\\
$N_\mathrm{H,i}$ (10$^{20}$ cm$^{-2}$) & $0.0^{+16.4}$ & $3.5^{+18.8}$\\
$kT_\mathrm{MCD}$/$kT_\mathrm{BB}$ (keV)& $0.118^{+ 0.016}_{-0.025}$  & $0.140^{+ 0.034}_{-0.037}$\\
$N_\mathrm{MCD}$/$N_\mathrm{BB}$ &$39^{+351}_{-21}$ & $23^{+433}_{-18}$\\
$L_\mathrm{abs}$ (10$^{43}$ erg~s$^{-1}$)\tablenotemark{a}  & $1.9^{+  0.6}_{ -0.6}$  & $  1.9^{+  0.9}_{ -0.7}$ \\
$L_\mathrm{unabs}$ (10$^{43}$ erg~s$^{-1}$)\tablenotemark{b} & $1.9^{+  4.4}_{ -0.5}$  & $2.7^{+  8.5}_{ -1.1}$ \\
$L_\mathrm{bol}$ (10$^{43}$ erg~s$^{-1}$)\tablenotemark{c} & $ 2.6^{+  7.8}_{ -0.8}$  & $6.1^{+ 31.0}_{ -3.0}$\\
$C/\nu$\tablenotemark{d} & $ 66.1( 78)$ & $ 66.3( 78)$\\
\hline
C1--C3 \\
Models & BB & MCD\\
$N_\mathrm{H,i}$ (10$^{20}$ cm$^{-2}$) & $0.0^{+15.8}$ & $0.0^{+16.0}$\\
$kT_\mathrm{MCD}$/$kT_\mathrm{BB}$ (keV)& $0.135^{+ 0.028}_{-0.024}$  & $0.169^{+ 0.043}_{-0.036}$\\
$N_\mathrm{MCD}$/$N_\mathrm{BB}$ &$4.6^{+18.8}_{-3.6}$ & $1.8^{+10.4}_{-1.4}$\\
$L_\mathrm{abs}$ (10$^{43}$ erg~s$^{-1}$)\tablenotemark{a}  & $0.4^{+  0.3}_{ -0.2}$  & $0.5^{+  0.4}_{ -0.3}$ \\
$L_\mathrm{unabs}$ (10$^{43}$ erg~s$^{-1}$)\tablenotemark{b} & $0.4^{+  0.6}_{ -0.2}$  & $0.5^{+  0.7}_{ -0.2}$ \\
$L_\mathrm{bol}$ (10$^{43}$ erg~s$^{-1}$)\tablenotemark{c} & $ 0.5^{+  0.8}_{ -0.2}$  & $1.0^{+  1.9}_{ -0.5}$\\
$C/\nu$\tablenotemark{d} & $ 36.6(34)$  & $ 34.8(34)$
\enddata 
\tablecomments{The C statistic was adopted, and the redshift of $z=0.41827$ was applied in all fits. All uncertainties are at the 90\% confidence level. \textsuperscript{a}Rest-frame 0.28--14.2 keV luminosity, corrected for Galactic absorption but not intrinsic absorption; \textsuperscript{b}rest-frame 0.28--14.2 keV luminosity, corrected for both Galactic and intrinsic absorption; \textsuperscript{c}the bolometric luminosity based on the total flux of the BB or MCD component; \textsuperscript{d}the C statistic and the degrees of freedom.}
\end{deluxetable}

\section{DISCUSSION}
\label{sec:discussion}
\subsection{The Wandering Massive BH Explanation}
Both the \textit{Chandra} and \textit{XMM-Newton} positions indicate
that our X-ray source XJ1417+52 could be associated with the galaxy
GJ1417+52, with an offset of $\sim$1\arcsec\ from the galaxy
nucleus. Therefore, it is an HLX candidate, with the peak luminosity
of $L_\mathrm{X}=3.8\times10^{43}$ erg~s$^{-1}$ in X1. The source
could be an accreting wandering BH of mass $\sim10^5$ \msun, assuming
that it was at the Eddington limit in X1. This mass is around the boundary
between IMBHs and SMBHs.  The temperature at the inner radius in a
standard thin disk at a given Eddington ratio is expected to depend on
the BH mass as $M_\mathrm{BH}^{-1/4}$, and Galactic BHBs tend to have
$kT_\mathrm{MCD}\sim1$ keV in the bright thermal state
\citep{remc2006,dogiku2007}. Therefore, our explanation of the source
as a BH of mass $\sim10^5$ \msun\ is supported by its very soft X-ray
spectra of $kT_\mathrm{MCD}\sim0.1$--0.2 keV.

Our source has a faint optical counterpart candidate sGJ1417+52. It
appears somewhat red in the optical and is thus unlikely to be the
emission from accretion activity.  It has a rest-frame absolute
$V$-band (close to the observer-frame F814W band) magnitude of
$\sim$$-15.9$ AB mag and is thus much more luminous than globular
clusters \citep[$M_V\gtrsim-13$ AB mag, refer to, e.g.,
][]{sijosa2007}. However, it is consistent with a compact dwarf
satellite galaxy, which is reminiscent of M32 in M31. It might have
been tidally stripped by GJ1417+52, resulting in the remnant
nucleus. A possible merging/interacting signature could be that the
outer/halo component of GJ1417+52 is a little twisted toward
sGJ1417+52, compared with the inner components (by $\sim$10$\degr$;
refer to the position angles of all components in
Table~\ref{tbl:hstfit}).

Tidally stripped galaxies are often used to explain UCDs and compact
elliptical galaxies (cEs) detected in nearby galaxies
\citep{hiinvi1999,drjogr2000,phdrgr2001,begoqu2010,nokafo2014,jerobr2015,chzo2015}. UCDs
have $R_\mathrm{e}\lesssim100$ pc, stellar mass $\lesssim10^8$ \msun,
and $M_V\gtrsim-14.0$ mag, while cEs have $R_\mathrm{e}\sim100$--$700$
pc, stellar mass $\sim10^8$--$10^{10}$ \msun, and $M_V$ from $\sim-14$
to $-20$ mag \citep[e.g.,][]{nokafo2014}.  With $R_\mathrm{e}$
$\lesssim$100 pc (or larger, considering possible systematic errors
due to, e.g., contaminating emission from GJ1417+52), the stellar mass
$\sim6\times10^7$ \msun\ (or larger if it contains old stellar
populations), and $M_V\sim-15.9$ AB mag, sGJ1417+52 is most likely a
massive UCD or a cE.

Due to the large distance, we cannot completely rule out that our
X-ray source is embedded in a globular cluster that is much smaller
than sGJ1417+52 and cannot be detected in the \textit{HST} images. We
have viewed sGJ1417+52 as the most likely, interesting counterpart to
our X-ray source, because its large size makes it more likely to host
a large BH of $\sim10^5$ \msun.

If XJ1417+52 is really a massive BH embedded in the remnant nucleus of
a dwarf satellite galaxy, one explanation for the outburst is the
tidal disruption of a surrounding star by the BH. Fundamental theory
predicts the mass accretion rate in TDEs to follow a
$(t-t_\mathrm{D})^{-5/3}$ decay, where $t_\mathrm{D}$ is the
disruption time \citep{re1988,re1990}. Because our source was only
detected in two epochs (2000 July and 2002 August), we cannot test
whether its luminosity decay followed this relation. In
Figure~\ref{fig:lumlc}, we plot a $t^{-5/3}$ decay curve that passes
through the X1 and C1--C3 data points and implies $t_\mathrm{D}$ to be
$\sim$11 months before X1. This decay curve predicts much lower fluxes
in the observations after C1--C3 than the detection limits, explaining
the non-detection of our source in those observations. Hydrodynamical
simulations predicted that the mass accretion rate in TDEs might decay
faster than $t^{-5/3}$ \citep{gura2013}. The disrupted star is more
likely to be a main-sequence star, instead of a white dwarf (WD). This
is because WD TDEs require smaller BHs ($\lesssim10^5$ \msun) for the
disruption to be outside the event horizon of the BH and should have
much shorter duration \citep[$\lesssim1$ yr,][]{rorahi2009} than our
event, which appeared to last for more than two years. We note that
the known candidate X-ray TDEs are all associated with the nuclei of
main galaxies, and our source could be the first X-ray TDE discovered
to be in the nucleus of a stripped satellite galaxy. One off-nuclear
optical TDE candidate has been reported in \citet{argasu2014}. Strong
narrow emission lines were not detected from sGJ1417+52 in our Gemini
observation, which could be because the lines as echoes of the X-ray
flare either had decayed significantly a decade after the disruption
or were too weak to be detected, as is often the case
\citep{gehako2003}.

Our source had a peak X-ray luminosity one order of magnitude higher
than that of \object{ESO 243-49 HLX-1} and two orders of magnitude
higher than those of other HLXs. Its distance is also one order of
magnitude larger than those of other HLXs (2.3 Gpc for our source
versus $<$200 Mpc for others). Therefore our source could be the most
luminous and the most distant HLX candidate ever discovered. It is the
only HLX candidate other than \object{ESO 243-49 HLX-1} showing very
soft X-ray spectra and with an early-type host (both are S0
galaxies). \object{ESO 243-49 HLX-1} also has an optical counterpart,
with a projected offset of 3.3 kpc from the nucleus of \object{ESO
  243-49}, thus similar to our source, but it appeared blue, unlike
the optical counterpart to our X-ray sources, which appeared
relatively red. The nature of the optical counterpart to \object{ESO
  243-49 HLX-1} is still somewhat under debate and could be a very
young ($\sim20$ Myr) stellar cluster with a mass of $\sim$$10^5$
\msun, with additional contribution from disk irradiation at long
wavelengths \citep{fasepf2012,fasegl2014}. The scenario of the remnant
nucleus of a bulgy or bulgeless satellite galaxy tidally stripped by
\object{ESO 243-49} has also been carefully explored through
$N$-body/smoothed particle hydrodynamics simulations
\citep{maanza2013a,maanza2013b}. The optical counterpart to our source
is much more massive (by about two orders of magnitude) and much older
(though still younger than typical red globular clusters) and is more
likely to be the remnant nucleus of a tidally stripped bulgy dwarf
galaxy. \object{ESO 243-49 HLX-1} showed quasi-periodic ($\sim$1 yr)
outbursts \citep{goloan2014}, thus unlikely due to complete tidal
disruption of a star in a single passage. Our source is consistent
with a transient due to a TDE, but with the sparse coverage of the
source, we cannot completely rule out the recurrent nature of the
source.

\subsection{Alternative Explanations}
\label{sec:dis_altexp}
Because of its close proximity to GJ1417+52, the relatively large
positional uncertainties of our X-ray source from \textit{Chandra} and
\textit{XMM-Newton} observations allow us to rule out that it is due
to the nuclear activity (either as a standard AGN or a TDE) in
GJ1417+52 only at the 95\% confidence level. However, there are other
arguments against the explanation as an AGN/TDE at the nucleus of
GJ1417+52. We did not identify XJ1417+52 as an AGN (either in GJ1417+52
or in the background) in \citet{liweba2012}, because none of the 753
AGNs in that study has X-ray spectra as soft as XJ1417+52. The large
long-term variability factor ($>$14) of XJ1417+52 found here is not
common in AGNs either; only 1.5\% of the 753 AGNs in \citet{liweba2012}
varied by factors of $>$10.  The AGN explanation is also disfavored
based on the lack of the [\ion{O}{3}] $\lambda$5007 in the Gemini
spectrum, which indicates little persistent nuclear activity in
GJ1417+52, at least two orders of magnitude lower than the peak X-ray
luminosity of our source. XJ1417+52 is unlikely a TDE at the nucleus of
GJ1417+52 because the central BH of this galaxy is probably too
massive ($\gtrsim$$10^8$\msun) to disrupt a solar-type star outside
the event horizon \citep{re1988}.

The high X-ray luminosities and ultrasoft X-ray spectra, which
probably lasted for $\gtrsim$2 years, make XJ1417+52 unlikely to be
the X-ray afterglow of a $\gamma$-ray burst (GRB) or a supernova (SN),
following similar arguments that we applied to a TDE candidate in
\citet{limair2015}. Essentially, X-ray spectra of the afterglows of
GRBs and SNs are generally hard, with $\Gamma_\mathrm{PL}\lesssim$2
\citep{im2007,lereme2013,grnove2013}. Although some ultralong GRBs
were discovered to exhibit very soft late-time X-ray spectra
\citep[e.g., ][]{pitrge2014,magula2015}, their hosts normally show
intensive star forming activity \citep[e.g.,][]{letast2014}, while
GJ1417+52 is an early-type galaxy. The long duration of XJ1417+52
cannot be explained with the prompt shock breakouts in SNe, which
could show soft X-ray spectra \citep{sobepa2008} but are expected to
be short \citep[less than hours,][]{nasa2012}.

\citet{liweba2012} did not identify XJ1417+52 as a coronally active
star because it has a 0.2--12 keV maximum flux to the $K$-band flux
ratio of $\log(F_\mathrm{X}/F_\mathrm{IR})>-0.65$ (the lower limit was
obtained because no counterpart was found in the 2MASS $K$ band),
higher than seen in coronally active stars
($\log(F_\mathrm{X}/F_\mathrm{IR})\lesssim-0.9$ in case of no
flares). The X-ray spectra of XJ1417+52 are much softer than seen in
stars too. The 0.2--0.5 keV to 0.5--1.0 keV hardness ratio is
$-0.55\pm0.04$ in X1, significantly lower than values of $\gtrsim$0.3
seen in stars \citep{liweba2012}.  With highly variable ultrasoft
X-ray spectra, XJ1417+52 is similar to supersoft X-ray sources (SSS),
most of which are due to nuclear burning of the hydrogen-rich matter
on the surface of a WD in the so-called close binary supersoft sources
or supersoft novae \citep{kava2006,gr2000}. However, such objects are
rare, with only a few tens found in our Galaxy \citep{kava2006,gr2000}
and the chance to find one within 1\arcsec\ of the center of a bright
galaxy should be very small. Besides, these objects have luminosities
typically $>10^{36}$ erg s$^{-1}$, while our source has a bolometric
luminosity of $6.3\times10^{30}$ erg s$^{-1}$, based on a BB fit
(redshift set to zero) and assuming a distance of 1 kpc. This
assumption on the distance is reasonable, given that our source is at
a high Galactic latitude of $60\degr$.

\section{Conclusions}
\label{sec:conclusions}
We have carried out a detailed study of the ultrasoft X-ray source
XJ1417+52, which is a candidate HLX in the S0 galaxy GJ1417+52 at
$z=0.41827$ ($d_L=2.3$ Gpc) in the EGS with a projected offset of
$\sim$1\farcs0 (5.2 kpc) from the nucleus. It was serendipitously
detected in five \textit{XMM-Newton} observations in 2000 July. Two of
them (X1 and X2) have enough counts for detailed spectral fits and
show very soft spectra that can be fitted with an MCD of
$kT_\mathrm{MCD}\sim0.13$ keV and $L_\mathrm{X}\sim4\times10^{43}$ erg
s$^{-1}$ in the source rest frame. It was still detected in three
\textit{Chandra} observations (C1--C3) in 2002 August, also exhibiting
ultrasoft spectra of $kT_\mathrm{MCD}\sim0.17$ keV but at a lower
luminosity of $L_\mathrm{X}\sim0.5\times10^{43}$ erg s$^{-1}$. The
source was not detected in later observations, with
$L_\mathrm{X}<0.9\times10^{43}$ erg~s$^{-1}$ in C4--C5 in 2005
October, $L_\mathrm{X}<1.1\times10^{43}$ erg~s$^{-1}$ in X3 in 2014
January, and $L_\mathrm{X}<0.25\times10^{43}$ erg~s$^{-1}$ in C6--C7
in 2014 September--October. Therefore the source has a long-term
variation factor of $>$14 and is likely a transient with an outburst
in 2000--2002. The source has a faint optical counterpart candidate
sGJ1417+52, which has $m_\mathrm{F606W}=26.33$ AB mag and
$m_\mathrm{F814W}=25.51$ AB mag in the observer frame in 2004
December, corresponding to the absolute $V$-band magnitude of
$\sim$$-15.9$ AB mag. All the properties of our source are consistent
with a massive BH of mass $\sim10^5$ \msun\ embedded in the remnant
nucleus of a satellite galaxy, with the outburst due to tidal
disruption of a surrounding star by the BH. Alternative explanations
such as a standard AGN in GJ1417+52 and Galactic SSSs are disfavored.

\acknowledgments We thank the referee for valuable comments that help improve the paper. Support for this work was provided by the National
Aeronautics and Space Administration through Chandra Award Number
GO4-15087X issued by the Chandra X-ray Observatory Center, which is
operated by the Smithsonian Astrophysical Observatory for and on
behalf of the National Aeronautics Space Administration under contract
NAS8-03060. This work is also supported in part by National Science
Foundation grant AST-1515084. We thank Chien Y. Peng for valuable
advice on the fitting of the galaxy with GALFIT and thank Jay Anderson
for providing aligned and bundled FLC frames.

Based on observations obtained with XMM-Newton, an ESA science mission
with instruments and contributions directly funded by ESA Member
States and NASA.

Based on observations obtained at the Gemini Observatory, which is
operated by the Association of Universities for Research in Astronomy,
Inc., under a cooperative agreement with the NSF on behalf of the
Gemini partnership: the National Science Foundation (United States),
the National Research Council (Canada), CONICYT (Chile), the
Australian Research Council (Australia), Minist\'{e}rio da
Ci\^{e}ncia, Tecnologia e Inova\c{c}\~{a}o (Brazil) and Ministerio de
Ciencia, Tecnolog\'{i}a e Innovaci\'{o}n Productiva (Argentina).

\appendix
\section{Astrometric Correction}
\label{sec:ascorr}
We needed to either align X-ray sources in various observations or
align X-ray sources and optical sources. We adopted a two-step method
to find the translation and rotation needed for the astrometric
correction, assuming that all sources in the reference frame and
sources to be aligned have positions and errors known.  The first step
is to find the translation and rotation that maximize the number of
matches $N_\mathrm{max}$ within the 99.73\% (i.e., $3\sigma$)
positional uncertainty. The second step is to find the translation and
rotation that minimize the reduced $\chi^2$ ($\chi$ is defined as the
separation of the matches divided by the total positional uncertainty)
for $fN_\mathrm{max}$ matches that have the smallest values of
$\chi$. Here $f$ represents the percentage of the matches used to
calculate the $\chi^2$ and is adopted to exclude possible spurious or
bad matches. Bad matches can occur if one or both of the matched
sources have bad positions for some reason, like being too close to
the CCD edge or being too close to other sources to be resolved well
by the detection tool. We have assumed $f=90\%$ throughout the paper.

The translation and rotation obtained for the astrometric correction
have uncertainties, increasing the positional uncertainties of the
aligned sources. We estimated the additional positional uncertainties
of the aligned sources associated with the astrometric correction
procedure, based on 200 simulations. In each simulation, we first
simulated the positions of the sources that have matches around the
positions of the matches in the reference frame, with positional
uncertainties following the combined positional uncertainties from the
reference frame and the frame to be aligned. Then the $\chi^2$
minimization for the $fN_\mathrm{max}$ matches that have the smallest
values of $\chi$ was repeated. The corrected positions for each source
from the simulations are then used to calculate the uncertainty
associated with the astrometric correction, which is added to the
original positional uncertainty of the source in quadrature.

\end{document}